\documentclass[10pt]{article}

\usepackage{epsfig}
\usepackage{a4wide}
\usepackage{amstex}

\begin{document}

\def\O{{\cal O}}
\def\N{{\cal N}}
\def\>t{>_{\scriptscriptstyle{\rm T}}}
\def\enu{\epsilon_\nu}
\def\pint{\int{\d^3p\over(2\pi)^3}}
\def\gint{\int[\D g]\P[g]}
\def\nbar{\overline n}
\def\barthe{{\bar\theta}}
\def\d{{\rm d}}
\def\e{{\bf e}}
\def\x{{\bf x}}
\def\y{{\bf y}}
\def\0x{\x^\smalze}
\def\sperpx{{x_\perp}}
\def\sperpk{{k_\perp}}
\def\sbperpk{{{\bf k}_\perp}}
\def\sbperpx{{{\bf x}_\perp}}
\def\perpx{{x_{\rm S}}}
\def\perpk{{k_{\rm S}}}
\def\bperpk{{{\bf k}_{\rm S}}}
\def\bperpx{{{\bf x}_{\rm S}}}
\def\p{{\bf p}}
\def\q{{\bf q}}
\def\zr{{\bf z}}
\def\R{{\bf R}}
\def\A{{\bf A}}
\def\v{{\bf v}}
\def\u{{\bf u}}
\def\U{{\bf U}}
\def\cm{{\rm cm}}
\def\l{{\bf l}}
\def\sec{{\rm sec}}
\def\Ckol{C_{Kol}}
\def\flux{\bar\epsilon}
\def\zq{{\zeta_q}}
\def\b{b_{kpq}}
\def\bun{b^{\scriptscriptstyle (1)}_{kpq}}
\def\bdu{b^{\scriptscriptstyle (2)}_{kpq}}
\def\z0q{{\zeta^{\scriptscriptstyle{0}}_q}}
\def\smalS{{\scriptscriptstyle {\rm S}}}
\def\smalze{{\scriptscriptstyle (0)}}
\def\smalI{{\scriptscriptstyle {\rm I}}}
\def\smalun{{\scriptscriptstyle (1)}}
\def\smaldu{{\scriptscriptstyle (2)}}
\def\smaltr{{\scriptscriptstyle (3)}}
\def\smalL{{\scriptscriptstyle{\rm L}}}
\def\smalD{{\scriptscriptstyle{\rm D}}}
\def\smal1n{{\scriptscriptstyle (1,n)}}
\def\smaln{{\scriptscriptstyle (n)}}
\def\smalA{{\scriptscriptstyle {\rm A}}}
\def\shell{{\tt S}}
\def\ball{{\tt B}}
\def\nav{\bar N}
\def\micron{\mu{\rm m}}
\font\brm=cmr10 at 24truept
\font\bfm=cmbx10 at 15truept
\centerline{\brm Three applications of scaling to}
\centerline{\brm inhomogeneous, anisotropic turbulence}
\vskip 20pt
\centerline{Piero Olla}
\vskip 5pt
\centerline{ISIAtA-CNR}
\centerline{Universit\'a di Lecce}
\centerline{73100 Lecce Italy}
\vskip 20pt

\centerline{\bf Abstract}
\vskip 5pt
The energy spectrum in three examples of inhomogeneous, anisotropic 
turbulence, namely, purely mechanical wall turbulence, the Bolgiano-Obukhov 
cascade and helical turbulence, is analyzed. As one could expect, 
simple dimensional reasoning leads to incorrect results and must be 
supplemented by informations on the dynamics. 
In the case of wall turbulence, an hypothesis of Kolmogorov cascade, starting 
locally from the gradients in the mean flow, produces an energy spectrum which 
obeys the standard $k^{-{5\over 3}}$ law only for $kx_3>1$, with $x_3$ the
distance from the wall, and an inverse power law for $kx_3<1$. An analysis 
of the energy budget for turbulence in stratified flows, shows the 
unrealizability of an asymptotic Bolgiano scaling. Simulation with a GOY model, 
leads instead to a $k^{-\alpha}$ spectrum for both temperature and velocity, 
with $\alpha\simeq 2$, and a cross-correlation between the two vanishing at 
large scales.  In the case of not reflection invariant 
turbulence, closure analysis suggests that a purely helical cascade, 
associated with a $k^{-{7\over 3}}$ energy spectrum cannot take place, 
unless external forcing terms are present at all scales in the Navier-Stokes 
equation.

\vskip 2cm
%\centerline{{\bf DRAFT} 25/7/97}
%\centerline{\it Submitted to}
\centerline{\it Phys. Rev. E}
\centerline{\it In press}
%\centerline{\it 28/7/97}
\vfill\eject
\centerline{\bf I. Introduction}
\vskip 5pt
Turbulence in nature is always inhomogeneous; the reason is the 
origin of the fluctuations, either in the instability of flow patterns, or 
in the presence in some finite volume, of temperature gradients, chemical 
reactions or external stirring. 
If the Reynolds number is large, however, there are turbulent fluctuations 
at scales much smaller than that of the forcing, and to them, the
idealization known as homogeneous isotropic turbulence can be applied. 

In practical applications, what one is interested in, is the effect of turbulence on
the mean flow and on transport, which is parametrized in terms of eddy 
viscosities and diffusivities (see e.g. \cite{pironneau}). In some cases, in 
order to calculate these quantities, some 
information on the turbulent energy spectrum is necessary, and Kolmogorov 
scaling \cite{k41} is usually assumed. For instance, in the 
derivation of Lagrangean diffusion models \cite{thomson87,borgas94}, the Kolmogorov 
scaling hypothesis is present 
explicitly through the assumption of Markovian velocity increments, at
time-scales below that of the energy containing eddies. 

What becomes necessary then, is some matching condition at the transition from the 
inertial range (small scale) to the energy containing range (forcing scale);
this is essentially the problem of connecting a region of 
$k^{-{5\over 3}}$ scaling, to the peak in the energy spectrum. Phenomenological 
theories have dealt with this problems \cite{tennekes};
more recently, the question of how fast the effect of inhomogeneity decays at small
scales has been put under exam both theoretically and using reduced models
\cite{grossmann94,yakhot94}.

There are situations, however, in which the problem of how far the inertial 
range preserves memory of the inhomogeneity of the forcing, becomes 
particularly serious. 

The most obvious way this can happen is when forcing takes place
at all scales. Notice that this does not necessarily require the presence,
to make an example, of obstacles of corresponding sizes in the flow.
Already in the case of wall turbulence \cite{tennekes}, one has mean flow 
gradients at lengths ranging from the viscous range to the height of the boundary 
layer, which leads to a situation of coexisting ''energy range'' and ''inertial range'' eddies, 
distributed at all scales.

An extended forcing range develops clearly, also in the presence
of stratification, due to the effect of buoyancy. In this case an additional
scale, the Obukhov length \cite{monin}, marking the transition from mainly 
mechanical 
to convection dominated turbulence, becomes important, and the way in which
mechanical and convective contributions to the dynamics balance one another, 
makes a description based on scaling rather non-trivial.

A third way, in which the small scale dynamics of turbulence could be modified 
by processes in the energy range, is when helicity is fed, together with
energy, into the system.
The importance of this process has been discussed recently by Yakhot 
\cite{yakhot94}, in the 
case of shear turbulence. Since the Navier-Stokes nonlinearity conserves both
helicity and energy, one wonders whether there could exist situations characterized
by an helicity cascade, analogous to the enstrophy cascade of two-dimensional 
turbulence \cite{kraichnan67,lesieur}.

All this neglects the presence of coherent structures and intermittency, which 
make an approach based on scaling and an hypothesis of homogeneous and 
isotropic inertial range questionable, even in the idealized case of spatially
homogeneous large scales \cite{k62,anselmet84,vincent91}. The importance of 
hairpin vortices in wall turbulence \cite{kline67,moin82} 
and even more, of plumes and effects at the boundary, in convective turbulence
\cite{krishnamurti86,castaing89,shraiman90} is well known. Helicity, on the other 
hand, has 
long been suspected to play an important role in triggering intermittency
in homogeneous turbulence \cite{levich83,benzi95}.

It should be mentioned, however, that with the exception of convective turbulence, 
intermittency and coherent structures seem to produce only minimal effects on 
energy spectra and transport coefficients. For this reason, they are not very 
interesting, when it comes to deriving turbulent models for engineering applications. 

The purpose of this paper is to study the behavior of the energy spectrum in 
the three examples of inhomogeneous turbulence listed. Dimensional analysis 
must necessarily be supplemented by information on the dynamics, to produce 
acceptable results. In wall turbulence, this will take the form of hypotheses
on the distribution of vortices and on the way they are generated. In the 
case of convective turbulence, an analysis of the energy budget in the Navier-Stokes 
and the temperature equations becomes necessary to verify the realizability
of different scaling hypotheses. In helical turbulence, the same task is
realized by means of closure analysis.

In the next section, wall turbulence is analyzed, assuming that at any height,
a Kolmogorov cascade is generated with integral scale equal to the height in exam. 
In section III, an analysis of the various possibilities for scaling in 
''homogeneous'' convective 
turbulence is carried on, using also results from simulations of a GOY model 
of the type introduced by Jensen et Al. \cite{jensen92}, plus buoyancy couplings. The scaling
predicted by Bolgiano \cite{bolgiano59,obukhov59,monin}, in particular, is taken under
exam.  Section IV 
is devoted to an analysis of helical turbulence using an EDQNM closure 
\cite{orszag,lesieur}. Section V contains the conclusions.
\vskip 20pt

\centerline{\bf II. Wall turbulence}
\vskip 5pt
To fix the ideas imagine a turbulent flow parallel to a horizontal plane,
characterized by a height $\delta$ and a stress at the surface (for unitary
fluid density) $v_*^2$.
If the Reynolds number $Re={v_*\delta\over\nu}$ is very large, with
$\nu$ the fluid viscosity, the mean velocity ${\bf V}$ will obey to a very good 
degree of approximation the logarithmic profile law: \cite{tennekes}
$$
V_1(x_3)={v_*\over\kappa}\log\Big({v_*x_3\over\beta\nu}\Big);\quad r_0\ll x_3\ll\delta,
\eqno(1)
$$
where $\kappa$ and $\beta$ are dimensionless constants depending on the roughness of the 
wall, $r_0=\nu/v_*$ is the flow inner length and $x_3$ is the distance from the 
wall. The law of the wall, Eqn. (1), does not provide informations about 
turbulent fluctuations, beyond what could be obtained from a mixing length 
approximation; indicating by ${\bf v}$ the turbulent velocities: 
$$
<v_1v_3>=-\nu_T{\partial V_1\over\partial x_3};\qquad \nu_T=lv_T,
\eqno(2)
$$
where the eddy size $l$, the characteristic turbulent velocity $v_T$ and the eddy viscosity 
$\nu_T$ all depend on the height $x_3$. In a scale invariant situation, using Eqns. (1) and 
(2):
$$
l(x_3)\sim x_3\quad{\rm and}\quad v_T\sim v_*,
\eqno(3)
$$
which means simply that gradients of strength $x_3^{-1}v_*$ at scale $x_3$ lead to vortices
of size $x_3$ and characteristic velocity $v_*$.
If one imagines that these vortices generate Kolmogorov cascades, spatially localized in 
height, some idea on the behavior of the energy spectrum can be obtained. 

This idea is not totally unreasonable, since, during the time it takes to the energy to be 
transferred to the viscous range, that is of the order of an integral time, eddies will move
at most by a distance of the order of an integral length $\sim x_3$.

Consider then the following picture (see Fig. 1): 
At a given height $x_3$, eddies of size $l_0\sim x_3$ are generated by instability of the mean 
flow. These ''mother eddies'' split into ''daughter eddies'' of size $l_{0\sigma}<l_0$ 
producing a Kolmogorov cascade; the index $\sigma$ indicates the point in the cascade
and is the logarithm of the ratio of the size of the daughter eddy to that of the original 
mother eddy. 
To these however, there will be superimposed mother eddies generated above $x_3$ and
which will have therefore size $l_\rho>l_0\sim x_3$. Also these produce cascades 
superimposed with the original one, with daughter eddies of size $l_{\rho\sigma}< l_\rho$.
Indicate:
$$
l_{\rho\sigma}=e^{-\sigma}l_\rho = e^{\rho-\sigma} x_3
\eqno(4)
$$
It is clear that $\sigma\ge 0$, but it is also true that $\rho\ge 0$. This
last condition means that, at height $x_3$, only mother eddies of size $l_\rho\ge x_3$ are 
present; smaller ones are generated at lower values of $x_3$. 

In the presence of a Kolmogorov cascade, one will have that the typical ratio of the 
velocity inside daughter and mother eddies will be proportional to 
$(l_{\rho\sigma}/l_\rho)^{1\over 3}\exp(-r/l_{\rho\sigma})$, with 
$r\simeq (x_3r_0^3)^{1\over 4}$ the viscous scale at height $x_3$.
One can then write for the velocity difference $\v_\l(\x)\equiv\v(\x+\l)-\v(\x)$:
%$$
%\v_\l(\x)\sim v_*\int_0^R\d\rho\int_0^\infty\d\sigma 
%\Big({l_{\rho\sigma}\over l_\rho}\Big)^{1\over 3}\exp(-r/l_{\rho\sigma})\int\d^3x'
%\u_\l(\x,\x',\rho,\sigma)
%\eqno(5)
%$$
$$
\v_\l(\x)\sim\sum_i\Big({l_{{\rho_i}{\sigma_i}}\over l_{\rho_i}}\Big)^{1\over 3}
\exp(-r/l_{{\rho_i}{\sigma_i}})
\u_\l(\x,\x_i',\rho_i,\sigma_i)
\eqno(5)
$$
where $\rho<R=\log(\delta/x_3)$, and $\u (\x,\x',\rho,\sigma)$  
is the normalized velocity at position $\x$, due to a vortex of type $\rho\sigma$ 
centered at $\x'$ ($\u_\l$ indicates finite difference with respect to the first argument).

\begin{figure}[hbtp]\centering
\centerline{
\psfig{figure=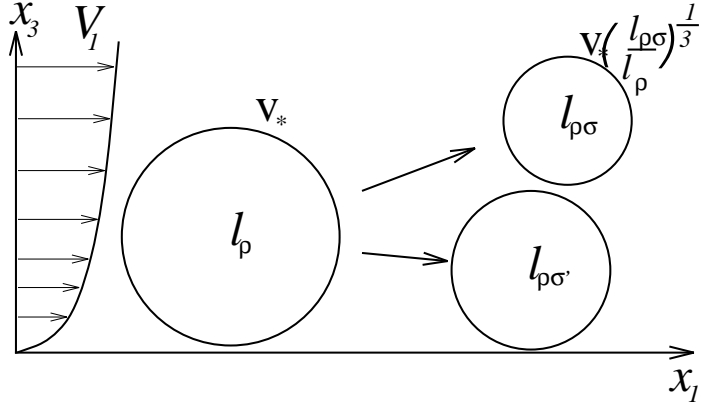,height=10.cm,angle=-90.}
}
\caption{Generation of Kolmogorov like cascade; for $\rho>0$, $l_\rho$ is larger
than the reference height $x_3$ at which the structure function $S(l,x_3)$ is being
measured.}
\end{figure}

An assumption on the statistics of $\u(\x,\x',\rho,\sigma)$ becomes therefore necessary. 
Spatial correlations in this model are assumed to arise from the finite extension
of individual eddies, while distinct eddies are taken to be uncorrelated. 
This leads to the expression:
$$
<u(\x,\x'\rho,\sigma)u(\y,\y',\rho',\sigma')>
=v_*^2\delta(\rho-\rho')\delta(\sigma-\sigma')\delta(\x'-\y')
C\Big({|\x-\y|\over l_{\rho\sigma}}\Big)
\eqno(6)
$$
with $C(l_{\rho\sigma}^{-1}|\x-\x'|)$ the normalized $[C(0)=1]$ auto-correlation for a single 
vortex. In general, vortices will be distributed with a density: 
$n(\x,\rho,\sigma)l_{\rho\sigma}^{-(3-\zeta_{\rho\sigma})}$, with $\zeta_{\rho\sigma}$ allowing for
the inclusion of intermittency corrections in the model. However, if 
$r_0\ll l_{\rho\sigma}\ll l_\rho\ll \delta$, scale invariance implies $n=n(l_\rho^{-1}x_3)$,
with $n(1)\sim 1$, while $\zeta=0$ for locally space filling (non-intermittent) turbulence.
Using Eqn. (6), the two-point structure function can be computed explicitly:
$$
S(l,x_3)=v_*^{-2}<v_l^2>\sim \int_0^R\d\rho\int_0^\infty\d\sigma 
F(\exp(\log(l/x_3)-\rho+\sigma))\exp\Big(-{2\over 3}\sigma
-\Big({r_0\over x_3}\Big)^{3\over 4}\e^{{3\over 4}(\sigma-\rho)}\Big)
\eqno(7)
$$
where $F(\alpha)=1-C(\alpha)$. Assuming smoothness of the velocity profile in an individual
vortex and finiteness of its total energy leads to the conditions on $F$:
$$
F(\alpha)=\O(\alpha^2)\quad (\alpha\ll 1)
$$
$$
\lim_{\alpha\to\infty}\alpha^2(F(\alpha)-1)=0.
\eqno(8)
$$
For $Re\to\infty$, one can then obtain asymptotic expressions for $S(l,x_3)$. 
For $l\ll x_3\ll\delta$, the integrals are dominated by the contribution at $\rho=0$ and 
$\sigma=\log(x_3/l)$, i.e. the contribution from the eddies down the cascade generated at $x_3$
which have size $l$. In this way, Kolmogorov scaling arises:
$$
S(l,x_3)=a\Big({l\over x_3}\Big)^{2\over 3}+\O\Big(\Big({l\over\delta}\Big)^{2\over3},
\Big({l\over x_3}\Big)^2\Big),\qquad l\ll x_3\ll\delta
\eqno(9a)
$$
For $l\gg x_3$ the dominant contribution is from mother eddies with size 
$x_3\le l_\rho<\min(l,\delta)$. Due to the uniform distribution in $\rho$, this leads
to a logarithmic form for the structure function:
$$
S(l,x_3)={3\over 2}\log(l/bx_3)+\O\Big(\Big({l\over\delta}\Big)^{2\over 3}\Big)\qquad 
x_3\ll l\ll\delta
\eqno(9b)
$$
$$
S(l,x_3)={3\over 2}\log(\delta/bx_3)+\epsilon(x_3/l)\qquad x_3\ll\delta\ll l
\eqno(9c)
$$ 
with $\epsilon(x_3/l)$ at most $\O((x_3/l)^2)$. The coefficients $a$ and $b$ and the various 
higher order terms in Eqns. (9a-c), all depend on the detailed shape of the function $F(\alpha)$. 
The one-dimensional energy spectrum \cite{note0} is:
$E_k(x_3)=\int_{-\infty}^{+\infty}(S(\infty,x_3)-S(l,x_3))\,\e^{ikl}\, \d l\sim 
\left|{\partial S(k^{-1},x_3)\over\partial k}\right|$ in the inertial range. Hence, 
one has  a standard 
$k^{-{5\over 3}}$ scaling for $l\ll x_3$ and a $k^{-1}$ scaling when $x_3\ll l\ll\delta$. 
A definite expression for the energy spectrum can be obtained fixing the form of the 
auto-correlation. Taking $C(\alpha)=\exp(-\alpha^2)$, one obtains: 
$$
E_k(x_3)=
\pi^{1\over 2}v_*^2x_3\int_0^R\d\rho\int_0^\infty\d\sigma
\exp\Big(\rho-{5\over 3}\sigma-{(kx_3)^2\over 4}\e^{2(\rho-\sigma)}
-\Big({r_0\over x_3}\Big)^{3\over 4}\e^{{3\over 4}(\sigma-\rho)}\Big)
\eqno(10)
$$
A plot of this spectrum for different values of $x_3/\delta$ is shown in Fig. 2. 
A fit of wind tunnel data, taken from \cite{naguib92}, is shown in Fig. 3; the data
correspond to values of the ratios: $x_3/\delta\simeq 0.01$ and 
$r_0/\delta\simeq 0.001$, i.e. to the experiment with $Re_{\theta}=7076$
and $y^+=28$ illustrated in Fig. 1 of that reference.

As discussed in \cite{naguib92}, $v_l$ samples in the range $l\gg x_3$, 
velocities corresponding to vortices generated much above $x_3$, while for $l\ll x_3$, 
it samples the daughter eddies of size $l$ generated in the cascade started at $x_3$.  
In \cite{perry86}, a $k^{-1}$ scaling was obtained by means of a hierarchy of hairpin 
vortices of prescribed shape, but the transition to the $k^{-{5\over 3}}$ range was left 
out of the description. This phenomenon is explained here as a natural consequence of  
implementing a Kolmogorov cascade in an inhomogeneous turbulence setting; in all cases,
hairpin vortices, and coherent structures in general, do not seem to be essential in 
obtaining such scaling behaviors.

\begin{figure}[hbtp]\centering
\centerline{
\input{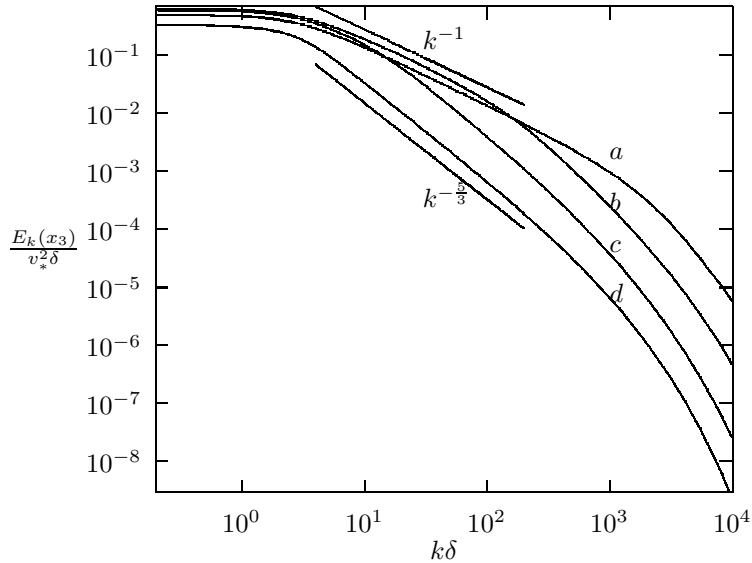}
}
\caption{One-dimensional energy spectra, for four different values of 
$x_3/\delta$. $a$: $x_3/\delta=0.001$; $b$: $x_3/\delta=0.01$; $c$: $x_3/\delta=0.1$; 
$d$: $x_3/\delta=0.5$. In all cases: $\delta/r_0\simeq 10^5$}
\end{figure}

The range $l\gg x_3$ is where turbulence ceases to be homogeneous and
isotropic. Notice that the divergence of $<v_1^2>$ as $x_3/\delta\to 0$ in Eqn. (9c)
forces an implicit assumption of anisotropy in the model, in order to avoid inconsistency 
with Eqns (2). Thus, although $<v_1^2>$ diverges as $x_3/\delta\to 0$, $<v_1v_3>$ 
remains equal to $v_*^2$; this result can be obtained assuming that 
the velocity of vortices generated much above $x_3$ is almost parallel to the wall
at $x_3$. Hence, while $<v_1^2>$ receives contributions from all vortices, up to 
size $\delta$, $<v_1v_3>$ receives contribution only from vortices up to size $x_3$.

\begin{figure}[hbtp]\centering
\centerline{
\input{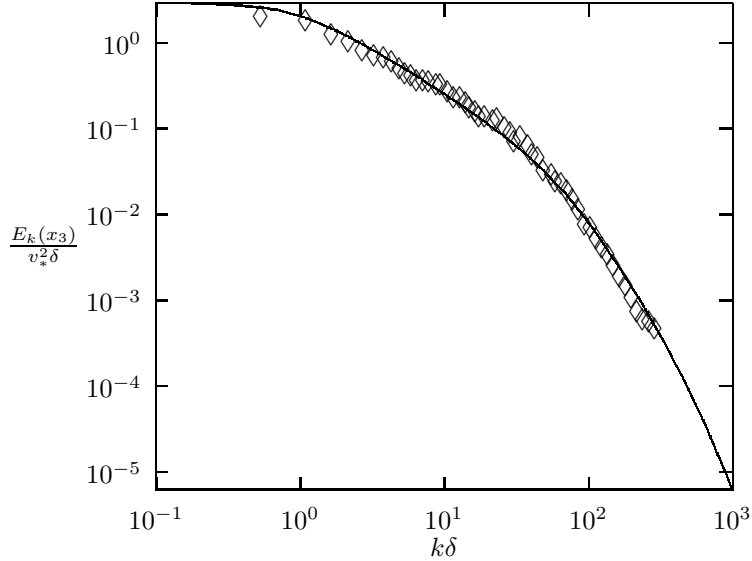}
}
\caption{Fit of wind tunnel data for $x_3/\delta\simeq 0.01$ and $r_0/\delta\simeq 0.001$.
The curve is shifted to the left to overlap with the data, corresponding to a shift
by $\simeq \log(3)$, of the limits of integration in $\rho$, in Eqns. (7) and (10).
Physically, this would correspond to a transition to anisotropy at $l\simeq 3x_3$.}
\end{figure}

Although essentially kinematic, this model is dynamically consistent. In spite of the
fact that all cascades overlap in space, the dominant 
interactions appear to be those among vortices of similar size, as in standard Kolmogorov 
theory, and belonging to the same cascade. A rough argument could be the following.
The relevant strain over an eddy of size $l$, from a cascade starting at height $x_3$, 
is produced by eddies of size $l'\ge l$. 
This strain will be of the order of: $v_*{l'}^{-1}(l'/x'_3)^{1/3}$ with $x'_3$ the height of 
generation of the second cascade. If $x'_3<x_3$, however, the volume in which the 
interaction takes place will be reduced by $x'_3/x_3$ and the effective strain will 
be of the order of $\min(1,x'_3/x_3)v_*{l'}^{-1}(l'/x'_3)^{1/3}$. The effective strain
will be maximum then for $x_3=x'_3$ and $l=l'$ corresponding to interaction with 
vortices of the same size and belonging to the same cascade.
\vskip 20pt

\centerline{\bf III. The case of ''infinite space'' convective turbulence}
\vskip 5pt
Stratification in a fluid produces buoyant forces which couple velocity and temperature
in the Navier-Stokes equation. In the limit of weak stratification in a large volume,
one can adopt the Boussinesq approximation \cite{tennekes}, in which the buoyant force 
in the Navier-Stokes equation, and the production term in the temperature equation, are
linearized respectively in the temperature and the vertical velocity fluctuation. For a 
unitary density medium:
$$
\Big({\partial\over\partial t}+\v\cdot\nabla\Big)\v=\nabla P+\nu\nabla^2\v
-g\Theta^{-1}\e_3\theta
\eqno(11a)
$$
$$
\Big({\partial\over\partial t}+\v\cdot\nabla\Big)\theta=\sigma\nabla^2\theta
+\Theta' v_3.
\eqno(11b)
$$
Here, $\sigma$ is the molecular diffusivity, $g$ is the gravitational acceleration,
$\theta$ and $\Theta$ are the
fluctuating and mean potential temperature, while $\Theta'\equiv {\d\Theta\over\d x_3}$.
The equations are linearly stable for $\Theta'>0$; in this case, external forcing is
necessary to achieve a stationary turbulent state. 

In homogeneous isotropic turbulence, one derives Kolmogorov scaling \cite{k41}, using dimensional
analysis with the quantities that are available: the scale $l$, the velocity difference
$v_l$ and the mean (kinetic) energy dissipation $\epsilon$; this leads to the well 
known result: $<v_l^2>\sim (\epsilon l)^{2\over 3}$. In the case of a stably stratified
medium, in the presence of mechanical forcing, Bolgiano \cite{bolgiano59,obukhov59} 
hypothesized an 
alternative situation, in which, potential energy transfer due to buoyancy forces, rather
than kinetic energy transfer, governs turbulence dynamics. Dimensional reasoning with 
the quantities available in Eqns. (11a-b), leads then to the scalings \cite{monin}:
$$
<v_l^2>= C_{vv}\Big({g\over\Theta}\Big)^{4\over 5}\epsilon_\theta^{2\over5} l^{6\over 5};\qquad
<v_l\theta_l>= C_{v\theta}\Big({g\over\Theta}\Big)^{1\over 5}\epsilon_\theta^{3\over 5} 
l^{4\over 5}; \qquad <\theta_l^2>= 
C_{\theta\theta}\Big({g\over\Theta}\Big)^{-{2\over 5}}\epsilon_\theta^{4\over 5} l^{2\over 5},
\eqno(12)
$$
where $\epsilon_\theta$ is the dissipation of temperature fluctuations:
$$
\epsilon_\theta=\sigma|\nabla\theta|^2.
\eqno(13)
$$
It is interesting to carry on this dimensional reasoning, directly inside Eqns. (11a-b).
After introducing the eddy turn-over frequency $\omega_l\sim l<v_l^2>^{1\over 2}$ and 
considering scales $l$ much larger than the dissipation lengths for $\v$ and $\theta$, we 
have from Eqn. (11a):
$$
\omega_l<v_l^2>\sim{g\over\Theta}<v_l\theta_l>\quad{\rm and}\quad
\omega_l<v_l\theta_l>\sim{g\over\Theta}<\theta_l^2>,
\eqno(14)
$$
while, from Eqn. (11b):
$$
\omega_l<v_l\theta_l>{\mathop\sim^{\rm ?}}\ \Theta'<v_l^2>\quad{\rm and}\quad
\omega_l<\theta_l^2>\sim\ \epsilon_\theta\, {\mathop\sim^{\rm ?}}\
\Theta'<v_l\theta_l>.
\eqno(15)
$$
The  question marks in Eqn. (15) indicate places in which the relation between terms
is ambiguous. It appears that the ambiguity lies in the source term $\Theta' v_3$
in Eqn. (11b). One sees immediately that the scaling described in Eqn. (12)
is obtained when the term in $\Theta'$ is negligible in Eqns. (11b) and (15).
Introducing the Obukhov length:
$$
L_*=\epsilon_\theta^{1\over 2}{\Theta'}^{-{5\over 4}}\Big({\Theta\over g}\Big)^{1\over 4},
\eqno(16)
$$
one realizes from Eqns. (12) and (15) that the condition of negligible $\Theta'$ term is 
equivalent to $l\ll L_*$. Thus, one has a source of temperature fluctuations at $l\le L_*$,
whose energy is transferred to smaller scales by action of the convective term 
$\v\cdot\nabla\theta$. These fluctuations provide the forcing for the velocity,
through the buoyant term ${g\e_3\theta\over\Theta}$, in Eqn. (11a). 

In order for a cascade of this form be present, the 
transfer of kinetic energy must be negligible, or equivalently, the frequency $\omega_l$
(which is the strain felt by eddies at scale $l$) must dominate the one that would be 
produced in a Kolmogorov cascade; indicating with $\epsilon$ the rate of kinetic energy 
production at the largest scales:
$$
\omega_l\sim v_{L_*}L_*^{-{3\over 5}}l^{-{2\over 5}}
\gg\epsilon^{1\over 3} l^{-{2\over 3}}\sim v_{L_*}L_*^{-{1\over 3}}l^{-{2\over 3}}
\eqno(17)
$$ 
which implies $l\gg L_*$. Thus, the two conditions of negligible $\Theta'$ in Eqn. (15) and
dominant buoyant force in Eqn. (14) restrict the possibility of Bolgiano scaling at most
to a finite range around $L_*$. 

An alternative situation, in which kinetic energy flows towards large scales, has been considered
in \cite{brandenburg92}. In this case, the condition provided by Eqn. (17) is not necessary 
anymore; however, the condition of negligible $\Theta'$ still forces Bolgiano scaling, to 
the range $l<L_*$. Now, this is the range in which, the equation for the velocity decouples 
from that for the temperature, and it is difficult to see a mechanism whereby buoyancy could
modify the nonlinear interaction in such a way to invert the direction of the energy 
transfer. 
If one restricts to this range, and maintains a situations of forward energy transfer, the 
decoupling forces the temperature to be advected like a passive scalar, which results in the 
well known Kolmogorov-Corrsin scaling \cite{corrsin51}: 
$$
<v_l^2>= C_{vv}(\epsilon l)^{2\over 3};\qquad
<v_l\theta_l>= C_{v\theta}\epsilon_{\theta}^{1\over2}\epsilon^{1\over 6}l^{2\over 3}
\qquad <\theta_l^2>=
C_{\theta\theta}\epsilon_\theta\epsilon^{-{1\over 3}}l^{2\over 3},
\eqno(18)
$$
The only way to obtain a power law fluctuation spectra for $l>L_*$ remains then, that:
$$
<v_l\theta_l>\sim g^{-1}\Theta\epsilon\sim{\Theta'}^{-1}\epsilon_\theta
\eqno(19)
$$ 
This would lead again to a $l^{2\over 3}$ spectrum for $<v_l^2>$ and $<\theta_l^2>$; 
in this case, however, there would be a privileged scale, the Obukhov length, which 
would fix the amplitude of the cross-correlation $<v_l\theta_l>$:
$$
{<v_l\theta_l>\over (<v_l^2><\theta_l^2>)^{1\over 2}}\sim\Big({L_*\over l}\Big)^{2\over 3}
\eqno(20)
$$
This leads to the situation, of correlations between $v_l$ and $\theta_l$ being the strongest
at $l\sim L_*$ and decaying at larger scales.

From these observations, it is clear that scaling behaviors should not be expected 
for $l>L_*$. An interesting question is then how, a toy system simulating a large 
range of scales like a GOY model, would behave in the presence of buoyancy. GOY
models (see \cite{bohr} and references therein) present a cascade 
of energy along a linear chain of coupled ordinary differential equations for the 
complex variables $u_n$, 
which are the analog of the velocity of eddies at scales $l_n=2^{-n}$ in real turbulence. 
These models have attracted great attention, due to the coincidence of the intermittent
properties of the moments $<|u_n|^n>$ and those of the structure functions $<v_l^n>$
in real turbulence. Jensen et Al. \cite{jensen92} have derived a generalization to 
the case of a passive scalar advected by a turbulent velocity field. It is easy to 
include in their model the effect of buoyancy; the resulting set of equations reads
($k_n=l_n^{-1}$):
$$
(\partial_t-\nu k_n^2)u_n=ik_n(au_{n+1}u_{n+2}+bu_{n-1}u_{n+1}+cu_{n-1}u_{n-2})^*-\alpha T_n+f_n
$$
$$
(\partial_t-\sigma k_n^2)T_n=ik_n[e_n(u_{n-1}T_{n+1}+u_{n+1}T_{n-1})
$$
$$
+g(u_{n-2}T_{n-1}+u_{n-1}T_{n-2})+h(u_{n+1}T_{n+2}+u_{n+2}T_{n+1})]^*+\beta u_n
\eqno(21)
$$
for $n=1,2,...N$, where the parameters $a,b,c,e,g,h$ are given by:
$$
a=1;\quad b=c=g=-e=-h=-{1\over 2}; 
\eqno(22)
$$
and variables $u_m$ and $T_m$ in the nonlinearities are set identically equal to zero 
for $m<1$ and $m>N$. The choice here is opposite to that of \cite{brandenburg92}, in the sense 
that, it has been preferred not to tamper with the nonlinearities, and to leave them in the 
same form as without buoyancy. However, also in this case, some interesting results are
obtained.

Equations (21-22) have been integrated numerically for both stable and unstable conditions. 
In the stable case, a constant forcing $f_n=(1+i)*10^{-3}\delta_{n4}$ has been used to 
sustain fluctuations. 

In the unstable case no external forcing was present, and the fluctuations organized
in such a way to push the Obukhov scale $N_*=\log_2(\epsilon_T^{-{1\over 2}}\beta^{5\over 4}
\alpha^{-{1\over 4}})$ towards the lowest available shell; a Kolmogorov cascade
ensued in all cases (see Fig. 4).

\begin{figure}[hbtp]\centering
\centerline{
\input{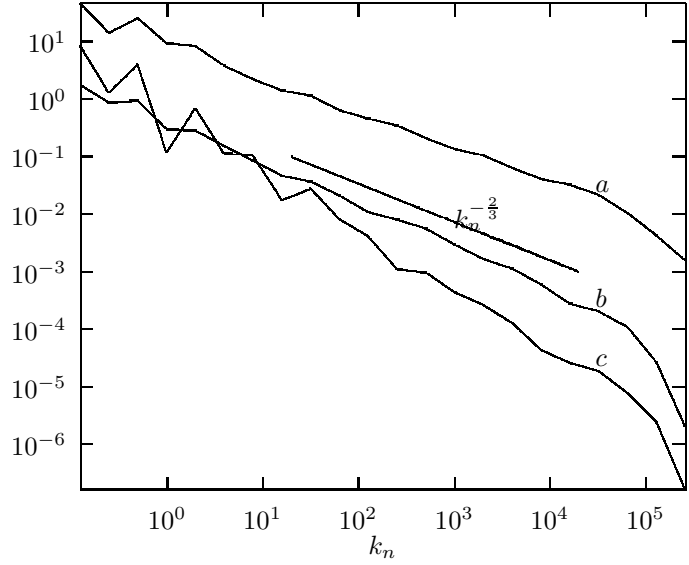}
}
\caption{GOY model simulation of convective turbulence under unstable conditions; values
of the parameters: $\alpha=0.01$ $\beta=-0.2$; $\nu=\sigma=10^{-7}$; no external forcing.
$a$: $<|T_n|^2>$; $b$: $<|u_n|^2>$; $c$: $<|u_nT_n|>$.
The buoyancy terms become of the same order of the others only for $k_n\le 1$;
}
\end{figure}

In the stable case, the presence of two additional parameters with which to play:
the forcing amplitude and wave number, allowed better control of $N_*$. When
$N_*$ was sufficiently large, neither Bolgiano scaling, nor the situation depicted
in Eqn. (18) took place, rather, a combination of the two, with steep $\sim k^{-1}$,
overlapping $T$- and $u$-spectra, and an almost constant cross-correlation $<u_nT_n>$
(see Fig. 5).

\begin{figure}[hbtp]\centering
\centerline{
\input{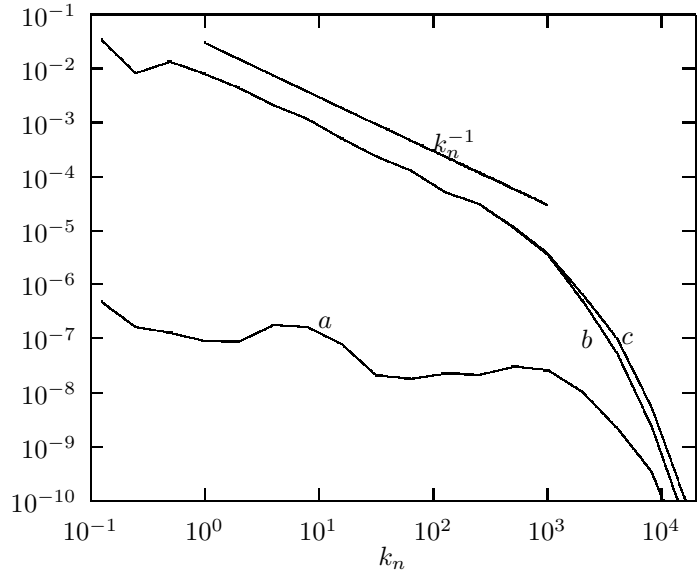}
}
\caption{GOY model simulation of convective turbulence under stable conditions; values
of the parameters: $\alpha=\beta=10.$; $\nu=\sigma=10^{-7}$, $f_n=10^{-3}(1+i)\delta_{n4}$.
$a$: $<|u_nT_n|>$; $b$: $<|u_n|^2>$; $c$: $<|T_n|^2>$.
The buoyancy terms are of the same order of the others, over the whole range of $k_n$,
down to the dissipation scale.
}
\end{figure}

An attempt has been carried also to generate a backward kinetic energy cascade, using 
small scales forcing (the forced shell was $n=20$). A physical interpretation of such 
a condition could be the presence of plumes generated elsewhere in the fluid, in some 
steep thermal layer. Using Eqns. (21-22), the outcome was, however, a $k^{-5/3}$ 
spectrum in all situations.

One can compare the behavior of $N_*$ in the two stability situations, looking at the
ratios of the
fluctuation amplitudes of the buoyancy terms $\alpha T_n$ and $\beta u_n$, to those
of $\partial_tu_n$ and $\partial_tT_n$. One sees in Fig. 6 how in the stable case,
buoyancy remains dominant over the whole inertial range, while the nonlinearity
dominates the dynamics in the unstable case.

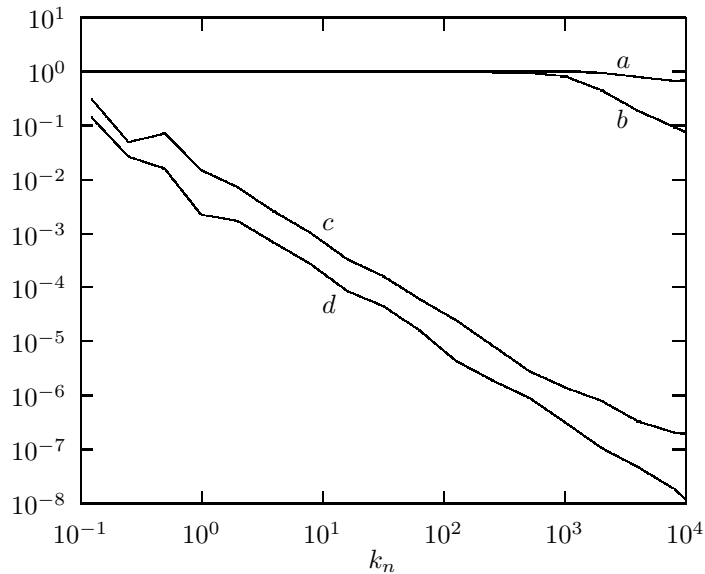
\begin{figure}[hbtp]\centering
\centerline{
% GNUPLOT: LaTeX picture
\setlength{\unitlength}{0.240900pt}
\ifx\plotpoint\undefined\newsavebox{\plotpoint}\fi
\sbox{\plotpoint}{\rule[-0.200pt]{0.400pt}{0.400pt}}%
\begin{picture}(1190,900)(0,0)
\font\gnuplot=cmr10 at 10pt
\gnuplot
\sbox{\plotpoint}{\rule[-0.200pt]{0.400pt}{0.400pt}}%
\put(176.0,877.0){\rule[-0.200pt]{4.818pt}{0.400pt}}
\put(154,877){\makebox(0,0)[r]{$10^1$}}
\put(1106.0,877.0){\rule[-0.200pt]{4.818pt}{0.400pt}}
\put(176.0,792.0){\rule[-0.200pt]{4.818pt}{0.400pt}}
\put(154,792){\makebox(0,0)[r]{$10^0$}}
\put(1106.0,792.0){\rule[-0.200pt]{4.818pt}{0.400pt}}
\put(176.0,707.0){\rule[-0.200pt]{4.818pt}{0.400pt}}
\put(154,707){\makebox(0,0)[r]{$10^{-1}$}}
\put(1106.0,707.0){\rule[-0.200pt]{4.818pt}{0.400pt}}
\put(176.0,622.0){\rule[-0.200pt]{4.818pt}{0.400pt}}
\put(154,622){\makebox(0,0)[r]{$10^{-2}$}}
\put(1106.0,622.0){\rule[-0.200pt]{4.818pt}{0.400pt}}
\put(176.0,537.0){\rule[-0.200pt]{4.818pt}{0.400pt}}
\put(154,537){\makebox(0,0)[r]{$10^{-3}$}}
\put(1106.0,537.0){\rule[-0.200pt]{4.818pt}{0.400pt}}
\put(176.0,453.0){\rule[-0.200pt]{4.818pt}{0.400pt}}
\put(154,453){\makebox(0,0)[r]{$10^{-4}$}}
\put(1106.0,453.0){\rule[-0.200pt]{4.818pt}{0.400pt}}
\put(176.0,368.0){\rule[-0.200pt]{4.818pt}{0.400pt}}
\put(154,368){\makebox(0,0)[r]{$10^{-5}$}}
\put(1106.0,368.0){\rule[-0.200pt]{4.818pt}{0.400pt}}
\put(176.0,283.0){\rule[-0.200pt]{4.818pt}{0.400pt}}
\put(154,283){\makebox(0,0)[r]{$10^{-6}$}}
\put(1106.0,283.0){\rule[-0.200pt]{4.818pt}{0.400pt}}
\put(176.0,198.0){\rule[-0.200pt]{4.818pt}{0.400pt}}
\put(154,198){\makebox(0,0)[r]{$10^{-7}$}}
\put(1106.0,198.0){\rule[-0.200pt]{4.818pt}{0.400pt}}
\put(176.0,113.0){\rule[-0.200pt]{4.818pt}{0.400pt}}
\put(154,113){\makebox(0,0)[r]{$10^{-8}$}}
\put(1106.0,113.0){\rule[-0.200pt]{4.818pt}{0.400pt}}
\put(176.0,113.0){\rule[-0.200pt]{0.400pt}{4.818pt}}
\put(176,68){\makebox(0,0){$10^{-1}$}}
\put(176.0,857.0){\rule[-0.200pt]{0.400pt}{4.818pt}}
\put(366.0,113.0){\rule[-0.200pt]{0.400pt}{4.818pt}}
\put(366,68){\makebox(0,0){$10^0$}}
\put(366.0,857.0){\rule[-0.200pt]{0.400pt}{4.818pt}}
\put(556.0,113.0){\rule[-0.200pt]{0.400pt}{4.818pt}}
\put(556,68){\makebox(0,0){$10^1$}}
\put(556.0,857.0){\rule[-0.200pt]{0.400pt}{4.818pt}}
\put(746.0,113.0){\rule[-0.200pt]{0.400pt}{4.818pt}}
\put(746,68){\makebox(0,0){$10^2$}}
\put(746.0,857.0){\rule[-0.200pt]{0.400pt}{4.818pt}}
\put(936.0,113.0){\rule[-0.200pt]{0.400pt}{4.818pt}}
\put(936,68){\makebox(0,0){$10^3$}}
\put(936.0,857.0){\rule[-0.200pt]{0.400pt}{4.818pt}}
\put(1126.0,113.0){\rule[-0.200pt]{0.400pt}{4.818pt}}
\put(1126,68){\makebox(0,0){$10^4$}}
\put(1126.0,857.0){\rule[-0.200pt]{0.400pt}{4.818pt}}
\put(176.0,113.0){\rule[-0.200pt]{228.855pt}{0.400pt}}
\put(1126.0,113.0){\rule[-0.200pt]{0.400pt}{184.048pt}}
\put(176.0,877.0){\rule[-0.200pt]{228.855pt}{0.400pt}}
\put(651,23){\makebox(0,0){$k_n$}}
\put(1018,807){\makebox(0,0)[l]{$a$}}
\put(1018,717){\makebox(0,0)[l]{$b$}}
\put(556,552){\makebox(0,0)[l]{$c$}}
\put(556,427){\makebox(0,0)[l]{$d$}}
\put(176.0,113.0){\rule[-0.200pt]{0.400pt}{184.048pt}}
\put(194,792){\usebox{\plotpoint}}
\put(938,790.17){\rule{11.500pt}{0.400pt}}
\multiput(938.00,791.17)(33.131,-2.000){2}{\rule{5.750pt}{0.400pt}}
\multiput(995.00,788.93)(4.306,-0.485){11}{\rule{3.357pt}{0.117pt}}
\multiput(995.00,789.17)(50.032,-7.000){2}{\rule{1.679pt}{0.400pt}}
\multiput(1052.00,781.93)(5.192,-0.482){9}{\rule{3.967pt}{0.116pt}}
\multiput(1052.00,782.17)(49.767,-6.000){2}{\rule{1.983pt}{0.400pt}}
\put(1110,776.67){\rule{3.854pt}{0.400pt}}
\multiput(1110.00,776.17)(8.000,1.000){2}{\rule{1.927pt}{0.400pt}}
\put(194.0,792.0){\rule[-0.200pt]{179.230pt}{0.400pt}}
\put(194,792){\usebox{\plotpoint}}
\put(709,790.67){\rule{13.731pt}{0.400pt}}
\multiput(709.00,791.17)(28.500,-1.000){2}{\rule{6.866pt}{0.400pt}}
\put(194.0,792.0){\rule[-0.200pt]{124.063pt}{0.400pt}}
\put(824,789.67){\rule{13.731pt}{0.400pt}}
\multiput(824.00,790.17)(28.500,-1.000){2}{\rule{6.866pt}{0.400pt}}
\multiput(881.00,788.93)(5.101,-0.482){9}{\rule{3.900pt}{0.116pt}}
\multiput(881.00,789.17)(48.905,-6.000){2}{\rule{1.950pt}{0.400pt}}
\multiput(938.00,782.92)(1.307,-0.496){41}{\rule{1.136pt}{0.120pt}}
\multiput(938.00,783.17)(54.641,-22.000){2}{\rule{0.568pt}{0.400pt}}
\multiput(995.00,760.92)(0.894,-0.497){61}{\rule{0.812pt}{0.120pt}}
\multiput(995.00,761.17)(55.314,-32.000){2}{\rule{0.406pt}{0.400pt}}
\multiput(1052.00,728.92)(1.123,-0.497){49}{\rule{0.992pt}{0.120pt}}
\multiput(1052.00,729.17)(55.940,-26.000){2}{\rule{0.496pt}{0.400pt}}
\multiput(1110.00,702.93)(1.179,-0.485){11}{\rule{1.014pt}{0.117pt}}
\multiput(1110.00,703.17)(13.895,-7.000){2}{\rule{0.507pt}{0.400pt}}
\put(766.0,791.0){\rule[-0.200pt]{13.972pt}{0.400pt}}
\put(194,749){\usebox{\plotpoint}}
\multiput(194.58,746.64)(0.499,-0.586){113}{\rule{0.120pt}{0.569pt}}
\multiput(193.17,747.82)(58.000,-66.819){2}{\rule{0.400pt}{0.284pt}}
\multiput(252.00,681.58)(2.076,0.494){25}{\rule{1.729pt}{0.119pt}}
\multiput(252.00,680.17)(53.412,14.000){2}{\rule{0.864pt}{0.400pt}}
\multiput(309.58,692.90)(0.499,-0.508){111}{\rule{0.120pt}{0.507pt}}
\multiput(308.17,693.95)(57.000,-56.948){2}{\rule{0.400pt}{0.254pt}}
\multiput(366.00,635.92)(1.062,-0.497){51}{\rule{0.944pt}{0.120pt}}
\multiput(366.00,636.17)(55.040,-27.000){2}{\rule{0.472pt}{0.400pt}}
\multiput(423.00,608.92)(0.751,-0.498){73}{\rule{0.700pt}{0.120pt}}
\multiput(423.00,609.17)(55.547,-38.000){2}{\rule{0.350pt}{0.400pt}}
\multiput(480.00,570.92)(0.856,-0.498){65}{\rule{0.782pt}{0.120pt}}
\multiput(480.00,571.17)(56.376,-34.000){2}{\rule{0.391pt}{0.400pt}}
\multiput(538.00,536.92)(0.696,-0.498){79}{\rule{0.656pt}{0.120pt}}
\multiput(538.00,537.17)(55.638,-41.000){2}{\rule{0.328pt}{0.400pt}}
\multiput(595.00,495.92)(1.062,-0.497){51}{\rule{0.944pt}{0.120pt}}
\multiput(595.00,496.17)(55.040,-27.000){2}{\rule{0.472pt}{0.400pt}}
\multiput(652.00,468.92)(0.793,-0.498){69}{\rule{0.733pt}{0.120pt}}
\multiput(652.00,469.17)(55.478,-36.000){2}{\rule{0.367pt}{0.400pt}}
\multiput(709.00,432.92)(0.866,-0.497){63}{\rule{0.791pt}{0.120pt}}
\multiput(709.00,433.17)(55.358,-33.000){2}{\rule{0.395pt}{0.400pt}}
\multiput(766.00,399.92)(0.726,-0.498){77}{\rule{0.680pt}{0.120pt}}
\multiput(766.00,400.17)(56.589,-40.000){2}{\rule{0.340pt}{0.400pt}}
\multiput(824.00,359.92)(0.713,-0.498){77}{\rule{0.670pt}{0.120pt}}
\multiput(824.00,360.17)(55.609,-40.000){2}{\rule{0.335pt}{0.400pt}}
\multiput(881.00,319.92)(1.103,-0.497){49}{\rule{0.977pt}{0.120pt}}
\multiput(881.00,320.17)(54.972,-26.000){2}{\rule{0.488pt}{0.400pt}}
\multiput(938.00,293.92)(1.370,-0.496){39}{\rule{1.186pt}{0.119pt}}
\multiput(938.00,294.17)(54.539,-21.000){2}{\rule{0.593pt}{0.400pt}}
\multiput(995.00,272.92)(0.894,-0.497){61}{\rule{0.812pt}{0.120pt}}
\multiput(995.00,273.17)(55.314,-32.000){2}{\rule{0.406pt}{0.400pt}}
\multiput(1052.00,240.92)(1.632,-0.495){33}{\rule{1.389pt}{0.119pt}}
\multiput(1052.00,241.17)(55.117,-18.000){2}{\rule{0.694pt}{0.400pt}}
\put(1110,222.67){\rule{3.854pt}{0.400pt}}
\multiput(1110.00,223.17)(8.000,-1.000){2}{\rule{1.927pt}{0.400pt}}
\put(194,720){\usebox{\plotpoint}}
\multiput(194.58,717.81)(0.499,-0.534){113}{\rule{0.120pt}{0.528pt}}
\multiput(193.17,718.90)(58.000,-60.905){2}{\rule{0.400pt}{0.264pt}}
\multiput(252.00,656.92)(1.518,-0.495){35}{\rule{1.300pt}{0.119pt}}
\multiput(252.00,657.17)(54.302,-19.000){2}{\rule{0.650pt}{0.400pt}}
\multiput(309.58,636.49)(0.499,-0.632){111}{\rule{0.120pt}{0.605pt}}
\multiput(308.17,637.74)(57.000,-70.744){2}{\rule{0.400pt}{0.303pt}}
\multiput(366.00,565.92)(2.945,-0.491){17}{\rule{2.380pt}{0.118pt}}
\multiput(366.00,566.17)(52.060,-10.000){2}{\rule{1.190pt}{0.400pt}}
\multiput(423.00,555.92)(0.841,-0.498){65}{\rule{0.771pt}{0.120pt}}
\multiput(423.00,556.17)(55.401,-34.000){2}{\rule{0.385pt}{0.400pt}}
\multiput(480.00,521.92)(0.856,-0.498){65}{\rule{0.782pt}{0.120pt}}
\multiput(480.00,522.17)(56.376,-34.000){2}{\rule{0.391pt}{0.400pt}}
\multiput(538.00,487.92)(0.679,-0.498){81}{\rule{0.643pt}{0.120pt}}
\multiput(538.00,488.17)(55.666,-42.000){2}{\rule{0.321pt}{0.400pt}}
\multiput(595.00,445.92)(1.196,-0.496){45}{\rule{1.050pt}{0.120pt}}
\multiput(595.00,446.17)(54.821,-24.000){2}{\rule{0.525pt}{0.400pt}}
\multiput(652.00,421.92)(0.751,-0.498){73}{\rule{0.700pt}{0.120pt}}
\multiput(652.00,422.17)(55.547,-38.000){2}{\rule{0.350pt}{0.400pt}}
\multiput(709.00,383.92)(0.594,-0.498){93}{\rule{0.575pt}{0.120pt}}
\multiput(709.00,384.17)(55.807,-48.000){2}{\rule{0.288pt}{0.400pt}}
\multiput(766.00,335.92)(0.939,-0.497){59}{\rule{0.848pt}{0.120pt}}
\multiput(766.00,336.17)(56.239,-31.000){2}{\rule{0.424pt}{0.400pt}}
\multiput(824.00,304.92)(1.062,-0.497){51}{\rule{0.944pt}{0.120pt}}
\multiput(824.00,305.17)(55.040,-27.000){2}{\rule{0.472pt}{0.400pt}}
\multiput(881.00,277.92)(0.732,-0.498){75}{\rule{0.685pt}{0.120pt}}
\multiput(881.00,278.17)(55.579,-39.000){2}{\rule{0.342pt}{0.400pt}}
\multiput(938.00,238.92)(0.713,-0.498){77}{\rule{0.670pt}{0.120pt}}
\multiput(938.00,239.17)(55.609,-40.000){2}{\rule{0.335pt}{0.400pt}}
\multiput(995.00,198.92)(0.954,-0.497){57}{\rule{0.860pt}{0.120pt}}
\multiput(995.00,199.17)(55.215,-30.000){2}{\rule{0.430pt}{0.400pt}}
\multiput(1052.00,168.92)(0.831,-0.498){67}{\rule{0.763pt}{0.120pt}}
\multiput(1052.00,169.17)(56.417,-35.000){2}{\rule{0.381pt}{0.400pt}}
\multiput(1110.00,133.92)(0.531,-0.494){27}{\rule{0.527pt}{0.119pt}}
\multiput(1110.00,134.17)(14.907,-15.000){2}{\rule{0.263pt}{0.400pt}}
\end{picture}
}
\caption{Plots of the ratios: $r_1={<|\alpha T_n|^2>\over <|\partial_tu_n|^2>}$ and 
$r_2={<|\beta u_n|^2>\over <|\partial_tT_n|^2>}$ vs. $k_n$. $a$: $r_1$; stable.
$b$: $r_2$; stable. $c$: $r_1$; unstable. $d$: $r_2$; unstable.
}
\end{figure}

In nature, of course, things go differently. First of all, the largest available scale 
$l_0$ corresponds to the size of the system. In the unstable case, steep thermal boundary 
layers develop rapidly and an approximation of constant temperature gradient ceases to
be applicable. Thus, the left portion of the spectra in Fig. 4 is not particularly 
meaningful, and the prediction that $L_*\to l_0$, under unstable conditions should 
not be trusted. In fact, in convective atmospheric turbulence, one has $L_*<l_0$, 
and a large scale, buoyancy dominated range, is indeed present \cite{kaimal72}.

Same amount of difficulties occur in the treatment of stable environments. 
In this case the problem is the idea of a purely large scale mechanical forcing. It 
is well known that, in these conditions, the large scale modes in turbulent shear 
layers are stabilized by buoyancy; thus, the peak in the forcing is moved to 
$l\sim L_*$, and a range like that of Fig. 4 is not generated \cite{kaimal72}.
However, a purely large scale forcing can be generated, if low frequency waves 
are present in the flows; in this case, a turbulent spectrum extending to the 
buoyancy dominated region $l< L_*$ becomes possible again (see e.g.  \cite{andreas87}).

\vskip 20pt
\centerline{\bf IV. Possibility of a purely helical cascade}
\vskip 5pt
The last situation that is taken into consideration is that of non-reflection invariant 
turbulence. It is well known that, beyond energy, the nonlinearity of the Navier-Stokes
equation has a second global invariant, the total helicity:
$$
H={1\over 2}\int\d^3 x\v(\x)\cdot[\nabla\times\v(\x)]
\eqno(23)
$$
In many ways, helicity is the counterpart in three dimensions of two-dimensional 
vorticity, and a natural question to ask is whether three-dimensional turbulence 
may exhibit multiple cascades, as it happens in two dimensions. The ability of
helicity to hinder energy transfer \cite{lumley67,lesieur}, in particular, suggests the 
possibility of an helicity cascade with no energy transfer, given appropriate
conditions on the forcing.

Helicity, however,
has the peculiarity of being a non-positive defined pseudo-scalar. Lack of positive
definiteness implies, in particular, that any triad of interacting modes, can exchange 
helicity in an arbitrary way, thus providing a source of helicity transfer
fluctuations.

This effect turns out to be important in GOY models; even in the case
of maximum injection of helicity, for a given energy injection rate, it can be shown
that the amount of the GOY equivalent of helicity: $H=\sum_n (-k_n)^n|u_n|^2$ \cite{benzi95},
which is produced by fluctuations, is much greater than the one coming from forcing
\cite{note}. Since
the variable $u_n$ in GOY models mimics in a surprising way the velocity inside individual
scale $l_n$ eddies, this may be a serious indication on the impossibility of an helicity 
cascade. 

Anyway, such a cascade seems impossible, also in a purely ''mean field'' description,
with an helicity transfer to small scales, assumed constant over the whole space.

The standard sequence of arguments,
leading to Kolmogorov scaling, can be carried on, assuming a constant helicity flux
$\epsilon_H$ to small scales; indicating with $H_l\sim l^{-1}v_l^2$, the content 
of helicity at scale $l$, one can then write:
$$
\epsilon_H\sim\omega_l H_l\sim l^{-2}v_l^3={\rm const.}\Longrightarrow
v_l\sim\epsilon_H^{1\over 3}l^{2\over 3}
\eqno(24)
$$
implying expressions for the energy and helicity spectra and for the eddy turn-over frequency:
$$
E_k=c_1\epsilon_H^{2\over 3}k^{-{7\over 3}};\qquad H_k=c_2\epsilon_H^{2\over 3}k^{-{5\over 3}};
\qquad\omega_k=c_3\epsilon_H^{1\over 3}k^{1\over 3}
\eqno(25)
$$
It is possible to obtain energy and helicity balance equations using statistical closure,
starting from the expression for the velocity correlation $U_{\bf k}^{ij}=
<|v^i_{\bf k}v^j_{-{\bf k}}|>$:
$$
2\pi U_{\bf k}^{ij}=k^{-2}P^{ij}({\bf k})E_k+k^{-4}\epsilon^{ijl}k^lH_k;\qquad
P^{ij}({\bf k})=\delta^{ij}-{k^ik^j\over k^2}.
\eqno(26)
$$
Lesieur has derived such balance equations within the EDQNM closure \cite{lesieur}. In this 
kind of closure \cite{orszag,lesieur}, the third order correlations $<vvv>$, which enter the 
equation for 
$U^{ij}_k$, are approximated by: $ <vvv>\simeq<v^\smalun v^\smalze v^\smalze>
+<v^\smalze v^\smalun v^\smalze> +<v^\smalze v^\smalze v^\smalun> $, with $v^\smalun$ 
the first order perturbative solution to 
a modified Navier-Stokes equation, with the viscous term $\nu k^2$  replaced 
by the eddy turn-over frequency $\omega_k$. The $v^\smalze$ are taken uncorrelated, so
that the resulting 4-point correlations split into products of 2-point correlations. 
Furthermore, the approximation: $<|v^i_{\bf k}(t)v^j_{-{\bf k}}(0)|>\simeq 
U_{\bf k}^{ij}\exp(-\omega_k|t|)$ is adopted.

The EDQNM equations for the energy and helicity balance read therefore 
\cite{lesieur}:
$$
\Big({\partial\over\partial t}+2\nu k^2\Big)E_k=
\int_{\Delta_k}\d p\d q\theta_{kpq}[{k\over pq}b_{kpq}E_q(k^2E_p-p^2E_k)
$$
$$
-{1\over p^2q}c_{kpq}H_q(k^2H_p-p^2H_k)]
\eqno(27a)
$$
and
$$
\Big({\partial\over\partial t}+2\nu k^2\Big)H_k=
\int_{\Delta_k}\d p\d q\theta_{kpq}[{k\over pq}b_{kpq}E_q(k^2H_p-p^2H_k)
$$
$$
-{k^2\over q}c_{kpq}H_q(k^2E_p-p^2E_k)],
\eqno(27b)
$$
where:
$$
\theta_{kpq}=(\omega_k+\omega_p+\omega_q)^{-1},\qquad b_{kpq}={p\over k}(xy+z^3)\quad
{\rm and}\quad c_{kpq}=z(1-y^2)
\eqno(28)
$$
with $\Delta_k$ the domain in which $k$, $p$ and $q$ can be the lengths of the 
sides of a triangle, and $x$, $y$ and $z$ the cosines of the angles opposite to 
these sides. 

The condition $\epsilon_H=const$ in Eqns. (24-25) and conservation of helicity
triad by triad guarantee that the spectra and frequencies of Eqn. (25) provide
a stationary solution for Eqn. (27b) for any value of the coefficients $c_i$.
The energy balance, which is given by Eqn. (27a) fixes instead, at stationarity,
the ratio of the two coefficients $c_1$ and $c_2$. Numerical integration of that 
equation leads then to the result:
$$
{c_2\over c_1}\simeq 3.316
\eqno(29)
$$
However, from the definition of helicity, one has:
$$
k^{-1}|H_k|=2\pi k<|[i{\bf k}\times\v_{\bf k}]\cdot\v_{-\bf k}|>\le
E_k=2\pi k^2<|v_{\bf k}|^2>
\eqno(30)
$$
which implies $|c_2|\le c_1$. Thus, Eqn. (29) cannot be satisfied, and an 
helicity cascade of the type described by Eqns. (24-25), does not seem 
to be possible.
\vskip 20pt
\centerline{\bf V. Conclusions}
\vskip 5pt
The aim of this paper was to obtain some information, on the effect of large
scale flow inhomogeneities on the form of the energy spectra in turbulent 
fluids. Some idealized situations have been studied by means of simplified
models and closure analysis.  The point in common in the three inhomogeneous 
turbulence situations considered, is that simple dimensional reasoning, 
either gives wrong answers, or does not lead to any answer at all. Of course 
this was something to be expected, and in a certain sense, there is nothing 
deep in this result. However, the practical consequences are important. 

The analysis carried on here clearly shows that a $k^{-1}$ range is a universal
feature of mechanical turbulent layers, which is independent of the presence of
coherent structures. 
If one is interested in diffusion in wall turbulence situations, a $k^{-1}$ range at 
scales $kx_3<1$ clearly makes a difference, with respect to a $k^{-{5\over 3}}$ 
spectrum, extending down to the inverse of the boundary layer thickness.  
Particles at distance $l>x_3$ will separate horizontally, in almost
a ballistic way: 
$l(t)\sim t$ (with logarithmic corrections), while Richardson law: 
$l_3(t)\sim t^{3\over 2}$ will dominate in the vertical direction. The modifications
that would be produced in dispersion models, for situations in which turbulence is 
predominantly mechanical, are clearly worth investigating.

The interest for the existence of Bolgiano scaling is more academic, although some
application to turbulent boundary layers in stable environments, in the presence of
forcing by low frequency waves, is possible \cite{andreas87}. This scaling 
has attracted some interest a few years ago to explain observations carried on in
liquid helium convection experiments \cite{procaccia89}. This approach has been 
criticized later by several authors \cite{shraiman90,lohse94}. The result of the 
analysis carried on here, suggests analogous difficulties for the existence of 
Bolgiano scaling in an idealized situation of convective turbulence in an infinite 
volume. The alternative however, which is characterized by velocity and temperature 
$k^{-{5\over 3}}$ spectra, has difficulties itself due the presence of a privileged
scale, the Obukhov length, dominating the dynamics, which weakens the very concept 
of an inertial range. GOY model simulations suggest indeed that neither scaling 
should be observed, rather, under stable conditions, a $k^{-2}$ spectrum for
both temperature and velocity, with correlation between the two, vanishing at
large scales, should develop. 

The possibility of helical turbulence has sparked recently some attention in
people interested in turbulence control \cite{levich97}. A $k^{-{7\over 3}}$ 
energy spectrum, associated with an helicity cascade, would imply
a decrease in the energy dissipation of the order of $Re^{-1}$ with respect to 
the standard $k^{-{5\over 3}}$ situation. (For equal total turbulent energy in 
the two cases: $\epsilon_H\sim\epsilon L^{-1}\quad\Longrightarrow\epsilon'\sim
l_d\epsilon_H$ with $\epsilon_H$ and $\epsilon'$ respectively the helicity and energy 
dissipation for a $k^{-{7\over 3}}$ situation, $\epsilon$ the energy dissipation 
for the corresponding $k^{-{5\over 3}}$, and $L$ and $l_d$ respectively the integral 
and viscous scales). The impossibility of an helicity cascade, suggested
by the EDQNM calculation carried on here, means simply that a $k^{-{7\over 3}}$ 
could not be obtained modifying the large scale forcing, and that action at all
scales (or alternatively at all frequency) would be necessary. Hence, turbulence
control by forcing the cascade to become helicity dominated, could be possible
only using a feedback system acting at inertial range frequencies.

\vskip 10pt
\noindent{\bf Aknowledgements}: I would like to thank Umberto Giostra and Federico Toschi
for interesting and helpful conversation. Part of this research was carried on at CRS4,
and I would like to thank Gianluigi Zanetti for hospitality.

\vskip 20pt
%\vfill\eject


\begin{thebibliography}{99}
\bibitem{note0} The separation $\l$ is taken implicitly in a fixed horizontal direction, in order
to allow simultaneous treatment of the two limits $l\gg x_3$ and $l\ll x_3$. The one-dimensional 
spectrum $E_k(x_3)$ corresponds, for $\l$ along $x_1$, to the energy spectrum of the signal
from a probe in the fluid.
\bibitem{note} Control over the ratio of helicity to energy injection in the simplest GOY 
model is achieved by forcing on two contiguous wave numbers $n$ and $n+1$, with
$f_n={aE_{n+1}u_n\over E_n+E_{n+1}}$ and $f_{n+1}={bE_nu_{n+1}\over E_n+E_{n+1}}$. The 
choice $b=a/2$ corresponds to no helicity input; $b=-a$ to no energy input.
\end{thebibliography}
\end{document}